On the problem of the Kondo-lattice model application to $CeB_6$.

N. E. Sluchanko*, A. V. Bogach*, V. V. Glushkov*, S. V. Demishev*,

M. I. Ignatov*, Yu. B. Paderno**, N. A. Samarin*, and N. Yu. Shitsevalova**

*Low Temperatures and Cryogenic Engineering Department, A.M.Prokhorov General Physics Institute of Russian Academy of Sciences, Vavilov street, 38, 119991 Moscow, Russia
** Department of Refractory Materials, Institute for Problems of Materials Science of Ukranian National Academy of Sciences, Krzhyzhanovsky street, 3, 03680 Kiev, Ukraine

Precision measurements of charge transport parameters (resistivity, Hall and Seebeck coefficients) have been carried out on high-quality single-crystals of cerium hexaboride in a wide temperature range 1.8-300 K. It is shown that in the temperature interval of 5 K < T < T* ~ 80 K the magnetic contribution in resistivity obeys the power law $\rho_m \sim T^{-1/\eta}$, which corresponds to the regime of weak localization of charge carriers with the critical index $1/\eta = 0.39 \pm 0.02$. In the same temperature interval an asymptotic behavior of thermopower $S \sim -lnT$ is found together with an essential decrease of the charge carriers mobility in $CeB_6$. A negative Hall coefficient anomaly has been detected at liquid helium temperatures. The data obtained are compared with the results predicted by the Kondo-lattice model and discussed also in terms of the theory of excitonic ferromagnetism.

**1.** It is generally believed that the cerium hexaboride $CeB_6$ is an archetypal example of a dense Kondo system [1, 2]. This compound crystallizes in the $CaB_6$-type structure that can be considered as a combination of two simple cubic lattices arranged from Ce-ions and $B_6$-octahedrons correspondingly and bound covalently to each other. A splitting $\Delta$ of the $Ce^{3+}$ $^2F_{5/2}$ state in the cubic crystalline field $\Delta = E(\Gamma_7) - E(\Gamma_8) \approx 530\ K$ [3] (see the insert in Fig. 1) exceeds considerably the Kondo temperature $T_K \approx 1-2\ K$ as estimated by different experimental methods (see, for example, [4]). According to the conclusion [5], one of the most important feature that gives rise to the anomalies of transport and thermodynamic characteristics in this analogue of monovalent metal with strong electron correlations is the coincidence between the number of itinerant electrons ($n_e$) and of the cerium 4f-sites ($n_{4f}$) - $n_{4f} \approx n_e$. It is basically assumed [5], that both an appearance of the complex magnetic structures in $CeB_6$ at liquid-helium temperatures and unconventional magnetic H-T phase diagram are associated with the coincidence of $n_{4f}$ and $n_e$ values, as well as with the competition between the Kondo scattering mechanism of charge carriers and the RKKY interaction of localized magnetic moments (LMM). In accordance with the result [6] for the dense Kondo systems, in the range of $n_{4f} \approx n_e$ it is natural to expect the largest amplitude of the Kondo maximum of resistivity $\rho(T)$. Indeed, the low temperature increase of resistivity by approximately a factor of 3 has been reported for $CeB_6$ at temperatures T ≤ 150 K [7, 8]. However, from the analysis of the experimental dependencies $\rho(T)$ itself and from the magnetic component in resistivity $\rho_m(T) = \rho(T) - \rho_{LaB6}(T)$ [7, 8] it is impossible to reveal a really extended section of the Kondo-like $\rho_m(T) \sim -lnT$ dependence in $CeB_6$.

One more problem in an approach to the interpretation of transport characteristics of $CeB_6$, which was noted by the authors of [4], consists in the fact that the Hall coefficient in cerium hexaboride, unlike the majority of so-called Ce-based Kondo lattices, is proved to be negative and, within the accuracy of measurements [4, 9], is independent of temperature and magnetic field in the interval 4.2—300 K. Thus, both the sign of the Hall coefficient and the temperature dependence $R_H(T)$ in $CeB_6$ are in contradiction with the predictions of the skew-scattering models [10-11] concerning the behavior of $R_H(T, H)$ in dense Kondo systems. Finally, there is another experimental fact associated with the behavior of the Seebeck coefficient $S(T)$ in $CeB_6$, that has not been adequately interpreted up to now. According to calculations [12] in the framework of the Kondo-lattice model, the broad positive maximum of $S(T)$ should be observed in the immediate neighborhood of $T_K$, whereas at $T_{inv} \approx 0.4T_K$ the sign of thermopower should be changed to negative. Since the Seebeck coefficient in $CeB_6$ acquires maximum at $T_{max}^S \approx 7—10\ K >> T_K \approx 1\ K$ [5, 13], there is the apparent discrepancy between the behavior of $S(T)$ and the theoretical results obtained in the Kondo-lattice model [12].

**2.** In such a situation, to elucidate an actual mechanisms responsible for the transport of charge carriers in cerium hexaboride, it is appropriate to carry out a precision measurements of the transport coefficients ($\rho$, $R_H$, and $S$) and to fulfill the comparative analysis of these results in a wide temperature range. With this purpose a high-quality single crystalline samples of $CeB_6$ were investigated. The synthesis technique of the cerium hexaboride single crystals and the characterization details are described in [14]. The measurements of $\rho(T)$, $R_H(T)$, and $S(T)$ were carried out in the temperature range of 1.8—300 K using the experimental setups similar to those earlier applied in [15, 16].

3. Figure 1 shows the $\rho(T)$ dependencies measured for the current aligned in various crystallographic directions in cerium hexaboride. The significant resistivity increase with lowering the temperature in the range T ≤ 150 K is changed by two stages of $\rho(T)$ decrease corresponding to antiferroquadropolar (AFQ) phase transition at $T_Q \approx 3.3$ K and antiferromagnetic (AFM) phase transition at $T_N \approx 2.3$ K (Fig. 1). As it is evidently deduced from the double logarithmic plot in Fig. 1, the $\rho(T)$ dependence in $CeB_6$ is well described by the power law $\rho(T) \sim T^{-1/\eta}$. This type asymptotic behavior usually corresponds to the regime of weak localization in the charge carriers' transport and it is established for $CeB_6$ in the entire temperature interval of 7—80 K. It is worth to mention that the observed value of the critical index $1/\eta \approx 0.39 \pm 0.02$ agrees well with the theoretical estimation - $1/\eta = 4/11$ found in [17]. The result $\eta = 11/4$ was obtained in [17] in the framework of the two-parametric scaling approach applied to the conducting system from the metallic side of the metal-insulator transition and with taking into account also the localization effects in combination with electron—electron interaction. It should be particularly emphasized that the increase in resistivity of $CeB_6$ within the temperature range 7 ÷ 80 K cannot be described by the logarithmic Kondo-like dependence $\rho \sim -\ln T$ neither for the initial curve $\rho(T)$, nor in analyzing the magnetic contribution in resistivity- $\rho_m(T) = \rho(T) - \rho_{LaB6}(T)$ (see curves 1 and 3 in Fig. 1).

Further we will concentrate on the results of high precision measurements of the Hall coefficient in $CeB_6$. Figure 2 shows the experimental dependence $R_H(T)$ (curve 1) and the parameter $\mu_H(T) = R_H(T)/\rho(T)$ (curve 2) which corresponds to the Hall mobility in the case of a system with one type of charge carriers. The data in Fig. 2 demonstrate obviously that the Hall coefficient of $CeB_6$ is negative and varies only slightly with the temperature in the interval of 5 ÷ 300 K. It should be mentioned also that the anomaly of the main component in $R_H(T)$ of negative sign is found for the first time in the neighborhood of the magnetic phase transitions at $T_Q$ and $T_N$ in $CeB_6$ (Fig. 2). Such a behavior of $R_H(T)$ in Kondo-lattice compound contradicts to the results of the skew scattering model calculations [10, 11] where the appearance of a broad maximum of the positive polarity Hall effect is expected in vicinity of the Kondo temperature $T_K(CeB_6) \approx 1 \div 2$ K.

One more feature that was detected in this study on the $R_H(T)$ dependence is a very smooth maximum near the liquid-nitrogen temperature (curve 1 in Fig. 2). From analysis of the mobility temperature dependence $\mu_H(T)$ also presented in Fig. 2, one can conclude evidently a qualitative changes of the scattering character of charge carriers in cerium hexaboride at T* ~ 80 K. When the temperature decreases in the interval 5K < T < T* ~80 K, the parameter $\mu_H(T)$ decreases also approximately by a factor of 3. It should be noted that, in Ce-based dense Kondo systems, the most ordinary situation is just opposite, i.e., $\mu_H(T)$ rises with decreasing the temperature (see, for example, [10,15]). Taking into account the value of lattice constant of $CeB_6$ $a \approx 4.14$ Å the data of $R_H(T)$ (Fig. 2) can be applied to estimate the absolute value and temperature dependence of the ratio $n_e/n_{4f} = 1/(eR_H n_{4f}) = f(T)$ in the approximation of one type of charge carriers. The result of the estimation is shown on the inset in Fig. 2.

When analyzing the results of precision measurements of the Seebeck coefficient in $CeB_6$ (curve 1 in Fig. 3), it should be emphasized that this parameter behavior is also affected by the above mentioned change in the character of scattering near *T* ~ *80 K*. In the temperature range where a weak localization type dependence of $\rho(T)$ is valid, thermopower increases sharply from the values of *S < 10 µV/K* (at *T > T* ~ 80 K*) which are typical for the metallic state to the values of *S ~ 70—90 µV/K* near the maximum of the *S(T)* dependence (Fig. 3). It is worth to point out also that very unusual logarithmic *S(T) ~ -lnT* dependence of thermopower is observed in the same temperature range where the resistivity follows to power law. Taking into account the additive character of a parameter $S/\rho = \Sigma S_i/\rho_i$, the discussed change in the regime of charge transport at *T ~ T* ~ 80K* can be most clearly demonstrated in the presentation *S/ρ =f(lgT)* (curve 2 in Fig. 3). As it may be detected from the data of Fig. 3, the behavior of the ratio (*S/ρ*) can be well described by the dependence $(S/\rho) \sim T^{1/\eta} lgT$ in the wide temperature interval *5K ≤ T ≤ T* ≈80 K* in $CeB_6$. It should be point out also that a noticeable deviations of the parameters *ρ, S*, and *S/ρ* from the analytical dependencies that were found in the interval Ib (Figs. 1 and 3) are observed at the temperatures below 5K. Evidently, the deviations can be interpreted in terms of an additional contributions to these transport characteristics appeared in the immediate neighborhood of the magnetic phase transitions at $T_Q$ and $T_N$ and, correspondingly, in AFQ and AFM magnetic phases of $CeB_6$.

**4.** The results of detailed measurements of charge carrier transport parameters (Figs. 1-3) obtained in this study for high-quality single crystals of cerium hexaboride were compared with the predictions in the frameworks of the approach based on the Kondo-lattice model. The analysis allows concluding that the traditional interpretation fails to explain the observed features of the transport characteristics of $CeB_6$.

When discussing an alternative approach to the explanation of observed anomalies, first of all, it is worth to consider the results of the electronic band structure calculations. From this point of view, a divalent hexaborides of alkaline- and rare-earth elements are semimetals [18], whereas the transition to a trivalent rare-earth ions in the $CaB_6$-type structure is accompanied by filling of the conduction band with an extremum at the X point in the Brillouin zone. According to the results [18-20] the conduction band has a significant dispersion and predominantly *5d*- character. Secondly, it is worth to note that the most consistent interpretation of anomalies of transport and thermodynamic characteristics both in the case of the doped divalent hexaborides (see, for example, [21, 22]) and for intermediate valence compound $SmB_6$ [23] is based on the appearance of exciton instability, which is accompanied by a partial or complete dielectrization of electronic structure (nesting, changes in the Fermi surface (FS) topology) in these materials. In this situation, the possibility of a similar scenario could not be ruled out for $CeB_6$, where a complicated FS structure has been found in [24-25]. The data of [24-25] suggest that the FS of $CeB_6$ consists of nearly spherical ellipsoids centered at the X points of the Brillouin zone with the necks in the [110] directions or with a small electron pockets. In accordance with the approach proposed earlier in [26, 27] and recently applied to hexaborides $La_xCa_{1-x}B_6$ and $La_xSr_{1-x}B_6$ [21-22], in the case of cerium hexaboride one would expect the development of exciton instability in *5d*-band at $T^* \sim 80K$. The phase transformation at $T^* \sim 80$ K as expected to be accompanied with a partial dielectrization of electronic spectrum within the possible transition to the charge density wave (CDW) state. In this scenario the spin density wave (SDW) transition into a phase of excitonic ferromagnet [26, 27], or into a spatially inhomogeneous magnetic multi-domain state with electronic phase separation [21, 22] may be expected at low temperatures $T \sim T_N$, $T_Q << T^*$. As a result, due to the excitonic instability and electronic phase separation at $T^*$, one can propose the appearance of a random potential in the samples of $CeB_6$ and, consequently, the weak localization asymptotic behavior in $\rho(T)$ (Fig. 1). In our opinion, this kind interpretation of the low temperature transport anomalies in $CeB_6$ can the strongly supported by the fact that the models explained the nature of magnetic phases are controversial and the character of phase transitions in this compound with a simple *bcc*- crystal structure still remains a subject of discussions [28]. Moreover, in $CeB_6$ the AFQ magnetic phase transition at $T_Q \approx 3.3K$ is characterized by a phase boundary $T_Q(H)$, which is identical to the behavior of the critical temperature $T_c$ vs magnetic field for the SDW state [29-30]. Moreover, it is known that the magnetic field improves the nesting properties of FS and may lead to the creation of SDW missing at H=0 [29,31-32].

Therefore, the excitonic ferromagnet approach should be considered as a very promising alternative to the Kondo-lattice model, which is conventionally applied for interpretation of the low temperature anomalies in $CeB_6$. At the same time, to elucidate the origin of the transition at T* ~ 80 K and extraordinary magnetic ground state of $CeB_6$, the comprehensive measurements of transport and magnetic properties of cerium hexaboride at liquid-helium temperatures are of a special interest. These investigations are in progress now and their results will be published elsewhere.

This work was supported by the Russian Foundation for Basic Research (project no. 04-02-16721), INTAS (no. 03-51-3036), the program "Strongly Correlated Electrons in Semiconductors, Metals, and Magnetic Materials" of RAS and the Russian Foundation for National Science Support.

FIGURE CAPTIONS

Fig. 1. Temperature dependence of resistivity $\rho(T)$: (1) $CeB_6$ for different current orientations, (2) $LaB_6$, and (3) magnetic contribution $\rho_m(T) = \rho(T) - \rho_{LaB6}(T)$ (see text). The logarithmic $(\rho \sim -\ln T)$ and exponential $(\rho \sim T^{-1/\eta})$ asymptotics are shown by dotted and solid lines respectively. The intervals I—III correspond to the paramagnetic (I(a, b)), AFQ (II) and AFM (III) phases in $CeB_6$. The inset shows the crystal field splitting of $Ce^{3+}$ $^2F_{5/2}$ state.

Fig. 2. Temperature dependencies of Hall coefficient for different orientations of electric current $I$ and external magnetic field $H$, together with the Hall mobility $\mu_H(T) = R_H(T)/\rho(T)$ in $CeB_6$. The inset shows the temperature dependence of the reduced concentration of charge carriers in $CeB_6$ (see the text).

Fig. 3. Temperature dependencies of Seebeck coefficient $S(T)$ and the ratio $S(T)/\rho(T)$ in $CeB_6$ (see the text). The intervals I—III correspond to the paramagnetic (I(a, b)), AFQ (II) and AFM (III) phases in $CeB_6$.

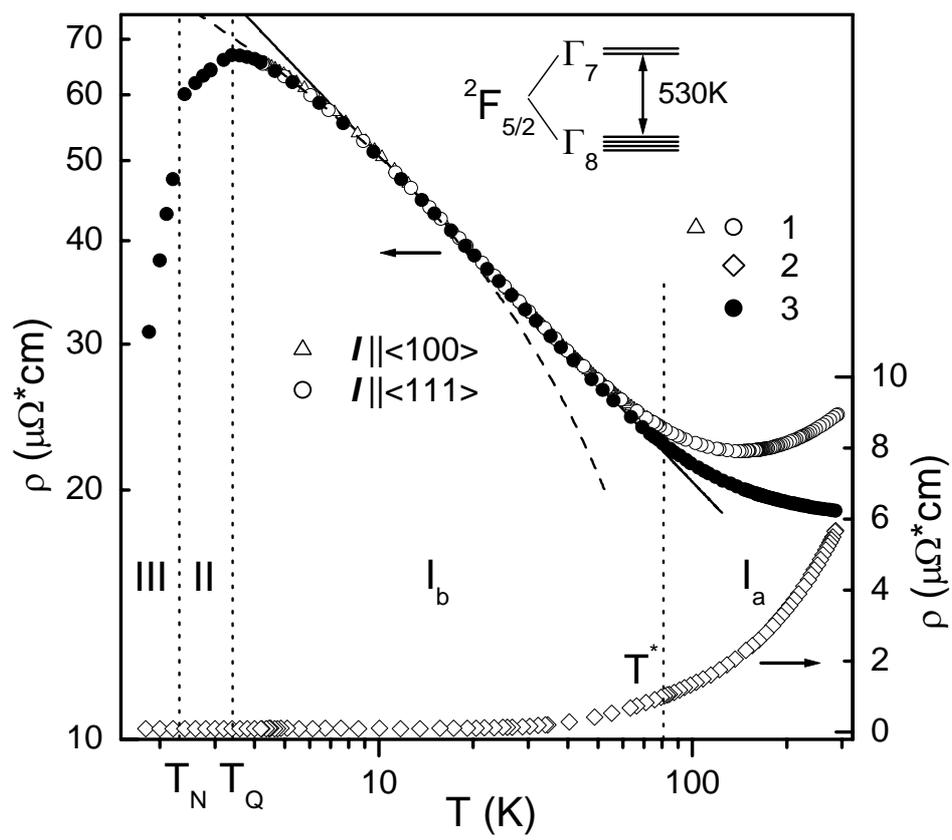

Fig. 1

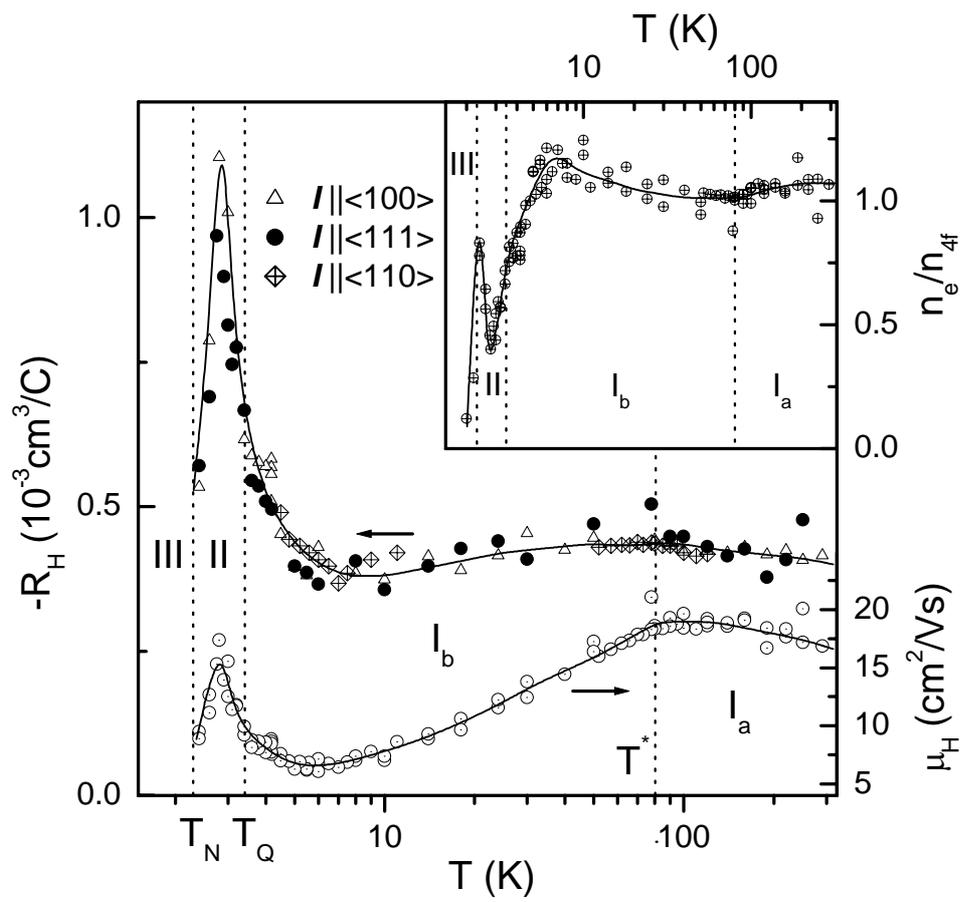

Fig. 2

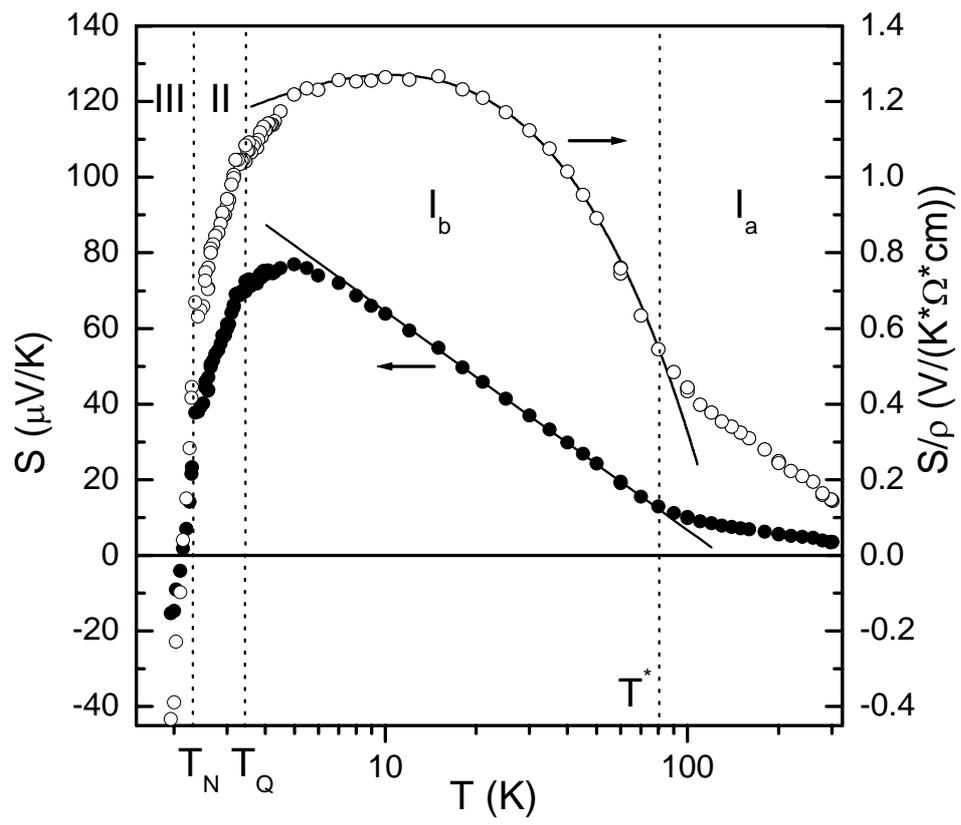

Fig. 3